# Nature of the bad metallic behavior of $Fe_{1.06}Te$ inferred from its evolution in the magnetic state

Ping-Hui Lin,[1] Y. Texier,[1] A. Taleb-Ibrahimi,[2] P. Le Fèvre,[2] F. Bertran,[2] E. Giannini,[3] M. Grioni,[4] and V. Brouet[1]

[1] *Laboratoire de Physique des Solides, Université Paris-Sud, UMR 8502, Bât. 510, 91405 Orsay, France*
[2] *Synchrotron SOLEIL, L'Orme des Merisiers, Saint-Aubin-BP 48, 91192 Gif sur Yvette, France*
[3] *Département de Physique de la Matière Condensée, Université de Genève, 24 Quai Ernest-Ansermet, 1211 Geneva, Switzerland*
[4] *Institut de Physique de la Matière Condensée, Ecole Polytechnique Fédérale de Lausanne, Station 3, CH-1005 Lausanne, Switzerland*

(Dated: June 11, 2013)

We investigate with angle resolved photoelectron spectroscopy the change of the Fermi Surface (FS) and the main bands from the paramagnetic (PM) state to the antiferromagnetic (AFM) occurring below 72 K in $Fe_{1.06}Te$. The evolution is completely different from that observed in iron-pnictides as nesting is absent. The AFM state is a rather good metal, in agreement with our magnetic band structure calculation. On the other hand, the PM state is very anomalous with a large pseudogap on the electron pocket that closes in the AFM state. We discuss this behavior in connection with spin fluctuations existing above the magnetic transition and the correlations predicted in the spin-freezing regime of the incoherent metallic state.

An interesting side product of the discovery of Fe-based superconductors is the exploration of correlation effects specifically linked with the orbital degeneracy. Hund's couplings may induce new types of correlations. They tend to align spins of the electrons in different orbitals and then favor the formation of local moments, which will interact with the itinerant electrons in a Kondo-like fashion [1]. Such a situation may result in very low coherence temperature scales, hence a large region of "bad metallic behaviors" appears [2], which was termed spin-freezing regime [3, 4]. $Fe_{1+y}Te$ (abbreviated as FeTe afterward) clearly qualifies as a "bad metal". Its resistivity increases with decreasing temperature down to the magnetostructural transition, $T_{ms}$=72 K [5]. In the PM phase, there is neither evidence for a Drude peak in the optical conductivity nor for a clear gap at low frequencies [5, 6]. The susceptibility in the PM state is Curie-like, suggesting localized moments rather than itinerant electrons [5]. In contrast, it is a much better metal in the AFM state. This difference may help to pinpoint the role of magnetic fluctuations in the bad metallic behavior. Moreover, it places correlation and magnetism at the heart of metallicity in one parent compound of Fe-based superconductors.

The relationship between superconductivity and magnetism remains one of the key questions in the field since the discovery of this new family of superconductors [7]. FeTe itself does not superconduct, but it does when doped with Se (up to 15 K at ambient pressure [8]). It exhibits a double stripe magnetic order [9, 10] (Fig. 1c) that contrasts with the stripe order observed in iron-pnictides [11] (Fig. 1b). The associated magnetic wavevector, $Q_{AFM} = (\pi, 0)$, does not nest the FS. This rules out the initial idea that magnetism in this family can be described as a simple nesting induced spin density wave (SDW) [12]. Indeed, the emerging picture is that the magnetic moments are formed due to local interactions on Fe and then order in a way that optimizes itinerancy [13]. A Dynamic Mean Field Theory (DMFT) approach successfully reproduces experimental values of the magnetic moments and relates them to the strength of correlations [14]. The largest value (2.2 $\mu_B$) is found in FeTe, which is also the most correlated. One qualitative reason is that the Te tetrahedra around the Fe is significantly distorted, pushing Te away from the Fe plane and reducing the overlap between the neighboring Fe [15]. Hence, FeTe provides a good opportunity to study magnetism in a "more localized" limit.

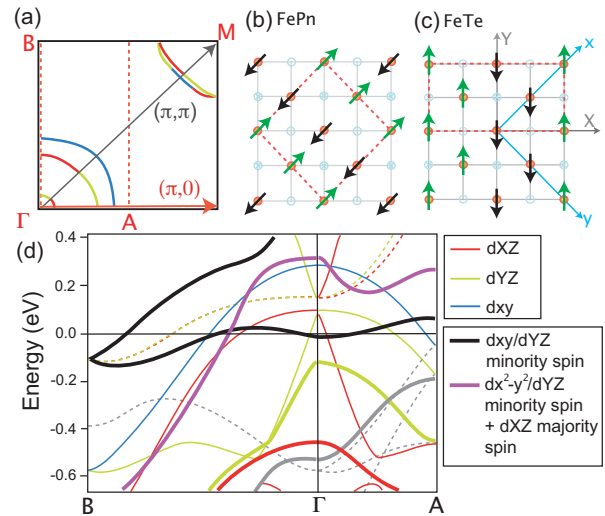

FIG. 1. (color online) (a) Sketch of the PM FS for FeTe in ¼ of the PM BZ (black lines). Red dotted lines indicate the AFM BZ. (b) Stripe magnetic order corresponding to $(\pi, \pi)$ observed in Fe-pnictides. (c) Double stripe magnetic order corresponding to $(\pi, 0)$ observed in FeTe. (d) Calculated electronic structure along the antiferromagnetic ($\Gamma A$) and ferromagnetic ($\Gamma B$) directions. Thin lines are bands in the PM state, with colors corresponding to the main orbital characters. Dotted lines are the bands folded with respect to the AFM BZ boundaries. Thick lines indicate the result of the calculation in the magnetic state. We use capital letters for the Fe-Te directions and small letters for the Fe-Fe directions.

The AFM state of FeTe is rather unusual as it is metallic but not driven by nesting. To better apprehend the impact of the magnetic order on the electronic structure, we show in Fig. 1 density function theory (DFT) calculation performed in the PM (see also [16]) and AFM states. We use the Wien2K package [17] with the experimental structure of FeTe [10] and checked the robustness of the results against different exchange correlation functions (the results are shown here for the local spin density approximation). The calculation converges to a magnetic moment of 2.2 $\mu_B$ close to the experimental value, in agreement with previous calculations [15]. As expected from the absence of nesting, the bands folded with the AFM periodicity (dotted lines) do not cross the PM bands near the Fermi level ($E_F$). Consequently, the stabilization of the magnetic state is not due to opening of gaps at $E_F$, as in the traditional SDW picture, but to a complete reorganization of the electronic structure. Fig. 1d shows that the hole $d_{XZ}/d_{YZ}$ bands shift below $E_F$ by 200 meV for $d_{YZ}$ and 450 meV for $d_{XZ}$. These different shifts emphasize the differentiation of the two directions in the magnetic state. The bands remaining at $E_F$ are completely rehybridized to optimize new conduction channels along the ferromagnetic direction. Except for a small electron pocket near Γ, no band crosses $E_F$ in the AFM direction. This explains the anisotropy detected recently in detwinned crystals [18].

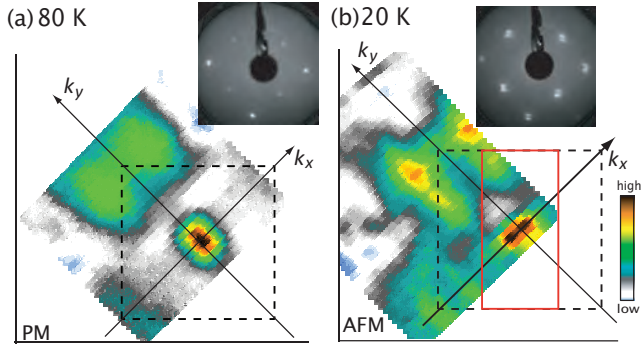

FIG. 2. (color online) FS map of FeTe at $k_z$=1 (estimated with $V_0$=12.5 eV [19]) obtained by integration of the spectral weight around $E_F$ over ±5 meV. It is measured at 70 eV with linear polarization along $k_x$ in the (a) PM state (80 K) and (b) AFM state (20 K). The red solid lines show the AFM BZ. The insets show the LEED at 100 eV of the ab plane in the corresponding state.

Angle-resolved photoelectron spectroscopy (ARPES) studies of Fe-chalcogenide have been mostly focused on $Fe_{1+y}Te_{1-x}Se_x$ [19, 20]. As for FeTe, Xia et al. [21] reported early on a FS at 27 K, but no study on the transition to the PM state. Quite different spectra, usually much broader, were then reported in ref. [22, 23]. Through the AFM transition, Zhang *et al.* [22], reported a large transfer of spectral weight and the appearance of a small coherent peak. However, the features were too unclear to address specific changes in band structure. Liu *et al.* discussed a similar "peak-dip-hump" lineshape appearing at Γ within a polaronic picture [23]. We clarify the connection between the electronic structure in PM and AFM phases by clearly identifying two hole bands and one electron band. These bands display a number of changes through the magnetic transition: some bands shift, exhibit larger Fermi velocities ($v_F$) and/or much clearer Fermi crossings below $T_{ms}$. These changes are in rough agreement with the prediction of our magnetic calculations. More importantly, they reveal a complete loss of intensity at $E_F$ along the electron pocket in the PM phase, which recovers in the AFM phase. This gives precise indications on how the metallic state is destroyed in the PM phase.

Single crystals of $Fe_{1+y}Te$ were grown by the Bridgman method [24]. The actual composition is determined by EDX analysis. The $T_{ms}$ of the samples, for which y=0.06, was determined from SQUID and transport measurements to be 72 K [25]. ARPES experiments were performed at the CASSIOPEE beamline at the SOLEIL synchrotron, with a Scienta R4000 analyser. The energy and angular resolution were ~25 meV and 0.2°. The samples were oriented by a 3-circle goniometer x-ray diffractometer prior to the experiment. FSs were acquired on a fresh surface, cleaved in ultra high vacuum at the temperature of the measurement. Temperature dependent measurements were carried out with both decreasing and increasing temperature, starting with samples cleaved at high and low temperatures, respectively. Results were found reproducible and reversible on more than 8 different samples. $E_F$ was measured regularly during the experiment on a Cu reference in electrical contact with the sample. The narrow spots observed in the LEED shown as insets in Fig. 2 demonstrate the good surface quality. At 20 K, the spots split due to the monoclinic twinning in the magnetic phase [10], confirming that the magnetic transition does take place at the surface, similarly as in the bulk (more details are given in [25]).

Fig. 2 evidences significant changes in the FeTe FS from the PM state at 80 K to the AFM state at 20 K. The hole pockets at the Brillouin zone (BZ) center seem to rotate by 90° in the AFM phase. The electron pockets at the BZ corners are difficult to see in the PM state but become clear and intense in the AFM state. They also seem to extend significantly towards the zone center. With the new AFM periodicity (red BZ), replica of hole and electron pockets are expected along the BZ edges. We indeed observe rather clear hole replicas in the dispersion, as in ref. [21, 22], but they have a weaker intensity and do not appear clearly in the FS. To understand the FS changes, it is necessary to look at the evolution of the different bands detailed in Fig. 3. These dispersions are measured along the diagonal of the PM BZ, where the twinned domains of our crystals give the same contribution.

In the PM state, the hole pockets are formed by 2 bands, shown in Fig. 3a1 and 3b1. They correspond to the calculated $d_{xz}$ and $d_{yz}$ bands. To compare with the calculation (white dotted line), we shifted the calculated bands down by 0.12 eV (meaning smaller hole pockets) and renormalized them by a factor 2. For the innermost band at Γ (Fig. 3a1), the intensity is concentrated at its top. In the calculation, it is degenerate with a small electron band, which actually approaches $E_F$ after the downward shift (as shown in Fig. 3a1) and may also give some contribution. We do not clearly observe the third hole band of $d_{xy}$ symmetry. As its contribution in ARPES is usually weak [19], it is difficult to conclude whether it is absent or just not clearly detected. Fig. 3c1 shows a shallow electron band with bottom near ~70 meV, which is most likely the $d_{xz}$ electron band. Its dispersion is quite flat and the intensity almost vanishes at $E_F$. To compare with calculation, we keep the same renormalization of 2 and shift down by 0.03 eV to fit the bottom of the band. We expect a different shift of hole and electron bands to conserve the number of carriers, but lack of information about $d_{xy}$, for both hole and electron pockets, does not allow more detailed investigation.

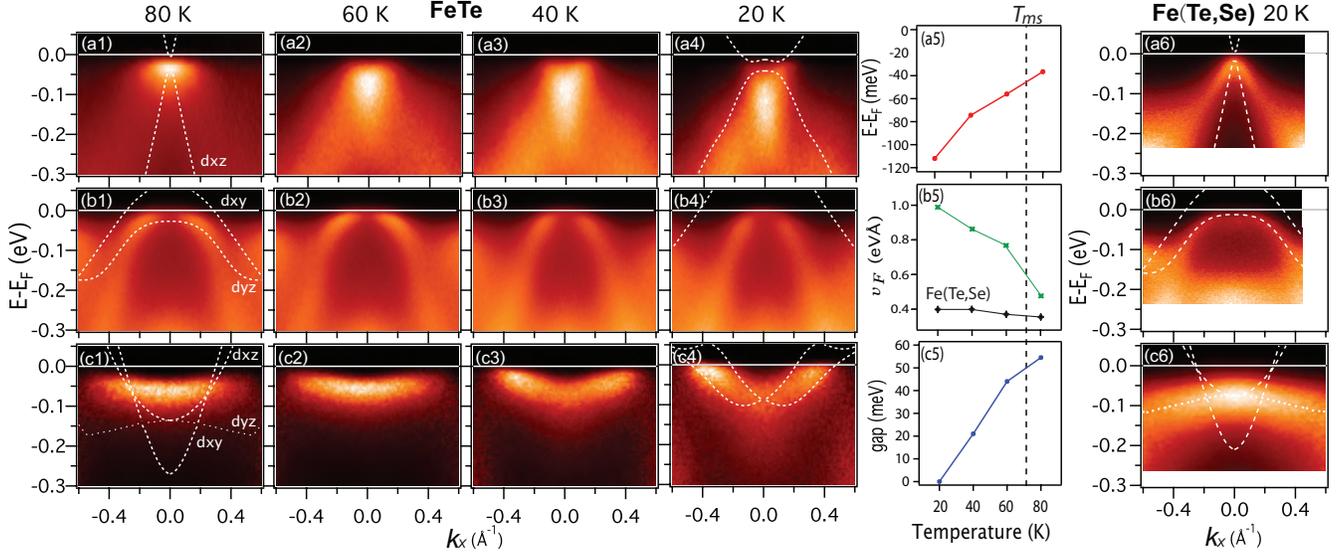

FIG. 3. (color online) Dispersion in FeTe (1-4) for indicated temperatures, measured at $k_z \cong 1$ (a) along $k_x$, at 38 eV with linear polarization (LP) along $k_x$, selecting orbitals even/$xz$, (b) along $k_x$, at 38 eV with LP in $yz$ plane (odd/$xz$) and (c) parallel to $k_x$, (0 corresponds to ($\pi$, $\pi$) in Fig. 2), at 70 eV with LP along $k_x$ (odd/$yz$). (a, b) were measured from 80 K to 20 K, while (c) was measured from 20 K to 80 K. (a5) Temperature dependence of the bottom of the inner hole band in (a1-a4). (b5) Temperature dependence of $v_F$ for the outer hole band in (b1-b4). That of $Fe_{1.09}Te_{0.78}Se_{0.22}$ is plotted for comparison as black points. (c5) Temperature dependence of the size of pseudogap at the electron pocket in (c1-c4). (a6, b6, c6) Dispersion in $Fe_{1.09}Te_{0.78}Se_{0.22}$ measured with the same experimental conditions as in FeTe. The band calculation in the PM or the AFM state are shown with white dashed line, with shift and renormalization described in the text. Only the bands allowed for the given polarization are shown.

When the temperature is lowered, the inner hole band shifts down by ~80 meV, as detailed in Fig. 3a5, revealing a small electron pocket on top of it (Fig. 3a1-a4, see also EDC stacks in supplementary information [25]). This brings strong intensity in the FS along that direction, explaining the apparent rotation of the hole pocket in the FS. This behavior is consistent with the expectation of the magnetic calculation, depicted as white dotted line with similar parameters as in the PM state (i.e., a down shift of 0.15 eV and renormalization by a factor 2). The same shift is reported in ref. [23], along with the appearance of a sharp peak at $E_F$ that is not as clear in our data, maybe due to slightly lower energy resolution.

The outer hole band does not shift as dramatically, but its shape near $E_F$ changes (Fig. 3b1-3b4 and [25]). At high T, a strong kink is observed near ~ 60 meV, which progressively disappears as the temperature decreases. The change of $v_F$ is summarized in Fig. 3b5 and compared to $Fe_{1.09}Te_{0.78}Se_{0.22}$, which shows no magnetic transition. In the calculation, this band has a larger $k_F$ (Fig. 3b4). It corresponds to the purple band in Fig. 1d, which is strongly rehybridized compared to the $d_{yz}/d_{xy}$ in Fig. 3b1. We speculate that this rehybridization takes place differently here, because $d_{xy}$ and $d_{xz}/d_{yz}$ are not correlated in the same way in the PM state [14].

For the electron band, there is a net evolution from the flat band at 80K to a "V" shape band at 20K, with clear Fermi crossing (Fig. 3c and [25]). In the magnetic calculation, this band changes shape and orbital character, but the main evolution here is due to the disappearance of the anomalous dispersion in the PM state, which, of course, cannot be described by a non-correlated calculation. To further investigate this anomalous behavior, we display in Fig. 4a the energy distribution curves (EDCs) at kF as a function of temperature. $k_F$ is defined as the point where the EDC is closest to $E_F$. At 20 K, there is a well defined peak and no gap. As the temperature increases, the peak moves gradually toward higher binding energy and the density at $E_F$ decreases, forming a sort of pseudogap. This peak position is reported in Fig. 3c5 and reaches 65 meV at 120 K. Fig.4a displays the symmetrized EDCs at various temperature to emphasize the pseudogap. Fig. 4b shows of the magnetic moment as a function of excess Fe contents[9, 10, 26]. While the resistivity of our sample is typical from a low quantity of excess Fe [25], it may be that some surface defects alter the way the magnetic moment develops at the surface. The particular evolution into the magnetic state measured by ARPES may be affected by such effects, but we believe it still gives a reliable picture of the global impact of magnetism on the electronic structure.

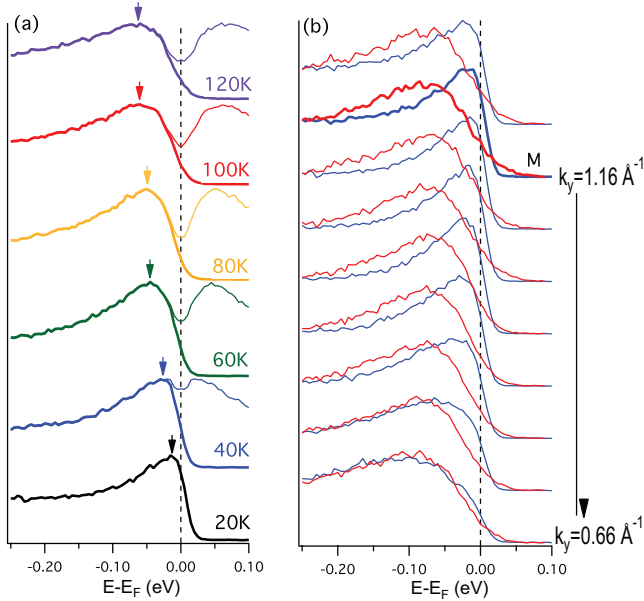

FIG. 4. (color online) (a) Thick Lines: EDCs at $k_F$ for the electron pocket at different temperatures. They are extracted from cuts along $k_x$ through $(\pi, \pi)$. Thin lines: Symmetrized EDCs to emphasize the formation of the pseudogap. (b) EDCs at $k_F$ for the electron pocket as function of $k_y$ at 20 K (blue line) and 80 K (red line). The thicklines represent the cuts through $(\pi, \pi)$.

These results suggest a bad metallic behavior for the PM state, since the $d_{xz}/d_{yz}$ hole pockets are very small, the $d_{xy}$ pocket is not observed and possibly completely incoherent and the electron pocket is largely gapped. The fact that the coherence of the large electron pocket is largely recovered in the AFM state proves that the loss of spectral weight is an intrinsic feature of the PM state. Indeed, our observations correlate well with transport measurements in FeTe. In contrast with $BaFe_2As_2$, the Hall coefficient $R_H$ remains large in the AFM state of FeTe [5]. This indicates that the FS does not break into very small pockets, as it does in $BaFe_2As_2$ due to the nesting induced FS reconstruction. Moreover, in FeTe, $R_H$ changes sign at $T_{ms}$, becomes negative in the AFM state[5], and the thermopower becomes much more negative [27]. These can be expected from our study with the larger coherence of the main electron pocket and the development of the small electron pocket at $\Gamma$.

Our study pinpoints the pseudogap on the electron pocket as the origin of the bad metallic state in FeTe. This was not resolved [25]. Strong band renormalization [14] and power law behavior of the imaginary part of the self energy [28] are expected to characterize the "spin-freezing" regime. Here, we use a modest value of renormalization of the order 2. In a Se-doped sample (Fig. 3a6, 3b6 and 3c6), we find comparable values, although significantly different correlations were predicted by theory [14]. In these systems where bands are far from linear and must be quite strongly shifted compared to calculations, these values have to be considered with care. Moreover, we might in fact reach a state, where the quasi-particles (QP) are so badly defined that speaking of renormalization is not meaningful anymore. This is certainly the case for the electron pockets, for which we fit the bottom of the band, but where there is no way to discuss the slope at $E_F$ due to the extremely small intensity. The loss of intensity through broadening and/or transfer of weight to incoherent structures is in fact the primary thing that can be expected in a bad metal. What is maybe unexpected in our findings is that the wipe out of the QP occurs in a well defined energy window (~50 meV, while the band bottom is broad but still well defined). In addition, it is k-dependent, as the intensity near $E_F$ is better defined for the hole pockets near $\Gamma$ than at the electron pocket. We note that the 50 meV energy window of the QP wipe out is a similar energy scale as that of the kink in the outer hole band and also to the dip in the "'peak-diphump'" analysis of ref. [23].

A way to particularize a region of the FS would be to consider fluctuations at some given wave vector. A natural choice here is the wave vector for spin fluctuations. It is known that despite the different $Q_{AFM}=(\pi, 0)$, magnetic fluctuations are present at $(\pi, \pi)$ in the PM state of FeTe and increase with Se doping [29]. The electron pocket in FeTe is large and does not nest well with the tiny $d_{xz}/d_{yz}$ hole pockets, but possibly better with the incoherent $d_{xy}$ hole pocket, if present (see Fig.1). This could explain the preferential appearance of a pseudogap on the electron pocket. Of course, the pseudogap should vanish when fluctuations disappear in the magnetic state of FeTe, as we observe. This idea is supported by the very curious shape of the electron pocket in $Fe_{1.09}Te_{0.78}Se_{0.22}$ in Fig. 3c6, which is barely detectable at $E_F$. This has not been explained to date. Although the intensity in Fig. 3c6 seems to follow the hole-like $d_{yz}$ band, this band is normally very effectively suppressed in these experimental conditions (as for Fig. 3c1). We suggest instead, by analogy with FeTe, that it is the electron band with an even larger pseudogap, associated with the enhanced $(\pi, \pi)$ fluctuations.

In conclusion, we have isolated the main bands of FeTe in the PM and AFM states. Both the FS and the bands are different from those reported by Xia et al. [21], but compare well with Se-doped FeTe [19, 20] and the studies of FeTe in ref. [22, 23]. The transition to the AFM state does not follow the traditional SDW scenario, as $Q_{AFM}=(\pi, 0)$, does not correspond to FS nesting. Our band structure calculations in the magnetic state give guidelines to interpret the evolution. Moreover, we find that the magnetic state is a good metal with no gap openings. In fact, it is the

PM state that deviates more strongly from the band calculation. In addition to the shifts and renormalizations usually observed in these systems, we observe broad features, kinks in the dispersion and a strong loss of weight near $E_F$ especially for the electron pockets. All these features point towards a bad metallic behavior and they disappear in the AFM state. This explains very well the evolution of transport measurements. We believe these data could help to understand how the bad metal is formed.

We thank S. Biermann, L. de'Medici and I. Paul for useful discussion, L. Petaccia for LEED and preliminary ARPES measurements on the BadElph beamline at ELETTRA, ZZ. Li and L. Fruchter for help with sample characterization. Financial support from the French RTRA Triangle de la physique and the ANR Pnictides" is acknowledged.

[1] K. Haule and G. Kotliar, New J. Phys., 11, 025021 (2009).
[2] L. de' Medici *et al.*, Phys. Rev. Lett., 107, 256401 (2011).
[3] P.Werner *et al.*, Phys. Rev. Lett., 101, 166405 (2008).
[4] M. Aichhorn *et al.*, Phys. Rev. B, 82, 064504 (2010).
[5] G. F. Chen *et al.*, Phys. Rev. B, 79, 140509 (2009).
[6] J. N. Hancock *et al.*, Phys. Rev. B, 82, 014523 (2010).
[7] J. Paglione and R. L. Greene, Nat. Phys., 6, 645 (2010).
[8] P. L. Paulose *et al.*, EPL, 90, 27011 (2010).
[9] W. Bao *et al.* Phys. Rev. Lett., 102, 247001 (2009).
[10] S. Li, C *et al.*, Phys. Rev. B, 79, 054503 (2009).
[11] Q. Huang *et al.*, Phys. Rev. Lett., 101, 257003 (2008).
[12] V. Cvetkovic and Z. Tesanovic, EPL, 85, 37002 (2009).
[13] M. D. Johannes and I. I. Mazin, Phys. Rev. B, 79, 220510 (2009)
[14] Z. P. Yin *et al.*, Nat. Mater., 10, 932 (2011).
[15] F. Ma *et al.*, Phys. Rev. Lett., 102, 177003 (2009).
[16] A. Subedi *et al.*, Phys. Rev. B, 78, 134514 (2008).
[17] P. Blaha *et al.*, Wien2K, An Augmented Plane Wave +Lacal Orbitals Program for Calculating Crystal Properties (Technische Universitat Wien, Austria, 2002).
[18] J. Jiang *et al.*, arXiv:1210.0397v1.
[19] A. Tamai *et al.*, Phys. Rev. Lett., 104, 097002 (2010).
[20] F. Chen *et al.*, Phys. Rev. B, 81, 014526 (2010).
[21] Y. Xia *et al.*, Phys. Rev. Lett., 103, 037002 (2009).
[22] Y. Zhang *et al.*, Phys. Rev. B, 82, 165113 (2010).
[23] Z. K. Liu *et al.*, Phys. Rev. Lett., 110, 037003 (2013).
[24] R. Viennois, *et al.*, J. Solid State Chem., 183, 769 (2010).
[25] See supplementary material.
[26] I. A. Zaliznyak, *et al.*, Phys. Rev. Lett., 107, 216403 (2011).
[27] I. Pallecchi *et al.*, Phys. Rev. B, 80, 214511 (2009).
[28] Z. P. Yin *et al.*, Phys. Rev. B, 86, 195141(2012).
[29] T. J. Liu *et al.*, Nat. Mater., 9, 716 (2010).

# 1. Sample Characterization:

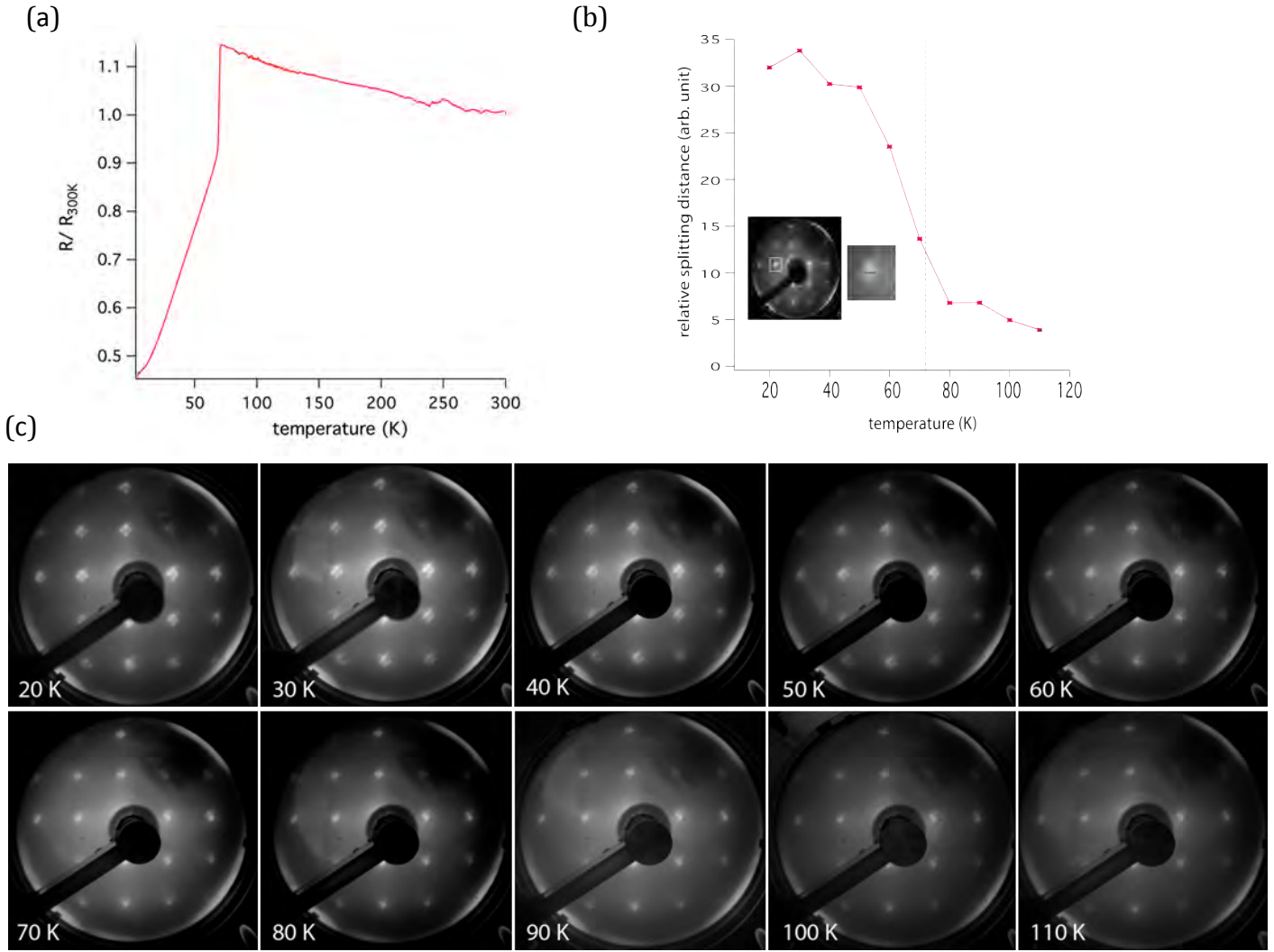

Sup. 1: (a) Temperature dependence of the resistance of FeTe normalized to value of 300 K. (b) Relative splitting distance of the LEED spot versus temperature. (c) Temperature dependent LEED pattern at 139 eV of the ab plane ($T_{ms}$= 72 K).

Sup. 1a displays the resistance of one sample used for the ARPES study. It displays a sharp drop in resistivity at $T_{ms}$=72 K, characteristic of the transition observed in the literature for samples with low contents of excess Fe.

To better characterize the transition closer to the surface, we performed LEED measurements as a function of temperature. The sharp LEED spots indicate the good quality of the sample surface (Sup. 1c, we note that some spots appear doubled at all temperatures. This can be due to 2 slightly different domains in the sample). At low temperatures, a clear splitting of the spots appears, evidencing the transition to monoclinic structure, which gives rise to twinning in the AFM phase. In Sup. 1b, we measure the distance between the LEED split spots, as shown in the inset. The splitting gradually appears below 70 K and saturates below 50 K. This evolution is similar to that of the magnetic moment (ref. [9,10]). We conclude that the transition appears similarly at the surface and in the bulk.

LEED patterns were acquired on the BL21B1 beamline, NSRRC, Taiwan.

## 2. Electron/hole pocket at Γ

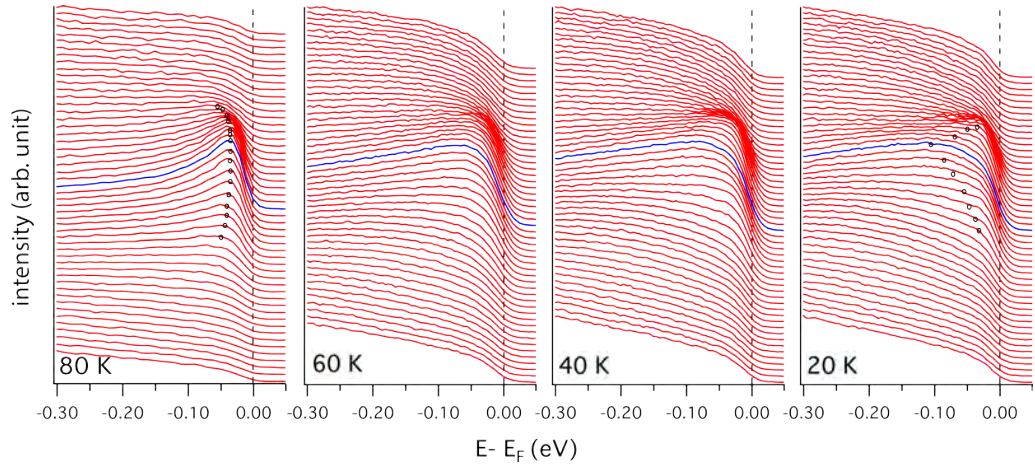

Sup. 2: EDC corresponding to Fig. (3a1-3a4). At 80 K, the EDCs indicate a hole like pocket slightly below $E_F$, while an electron like pocket develops as the temperature decreases. The blue line highlights the EDC at Γ. Its maximum is reported in Fig. 3a5, it shifts down with temperature in very good agreement with ref. [23].

## 3. Outer hole pocket at Γ

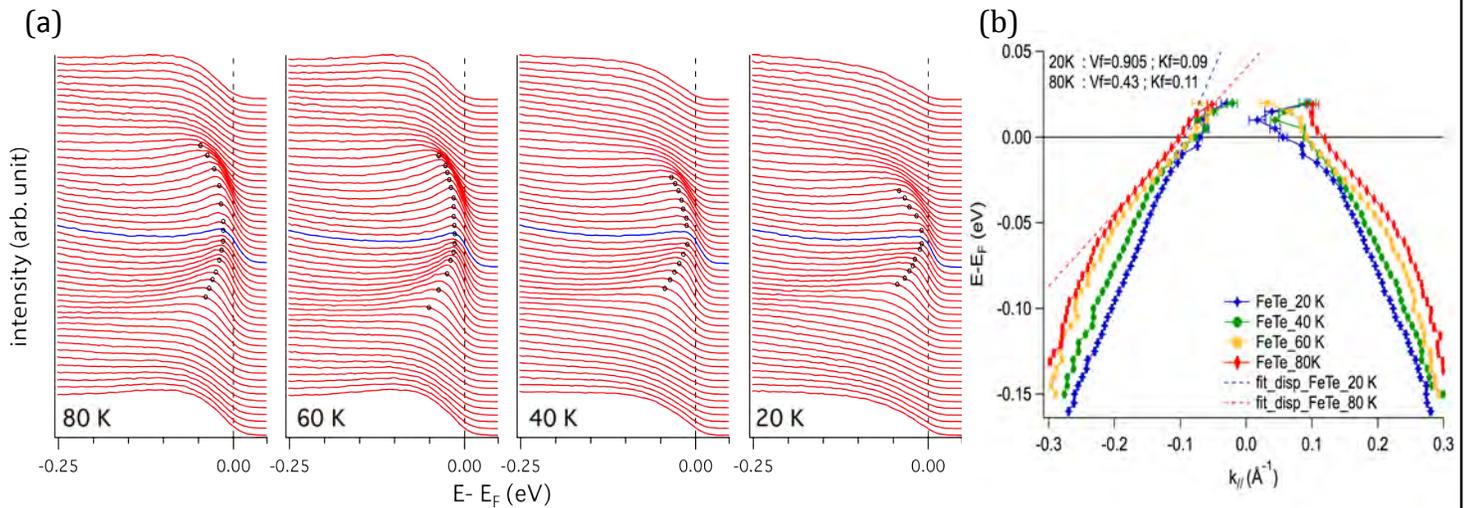

Sup. 3: (a) EDC corresponding to Fig. (3b1-3b4). Open circles mark the local maxima. Blue line: EDCs at the Γ point at various temperatures. (b) Temperature dependent dispersions extracted by MDC from Fig. (3b1-3b4). Fermi Velocities, $v_F$, for 80 K and 20 K are obtained by fitting the dispersion.

## 4. Electron pocket at M:

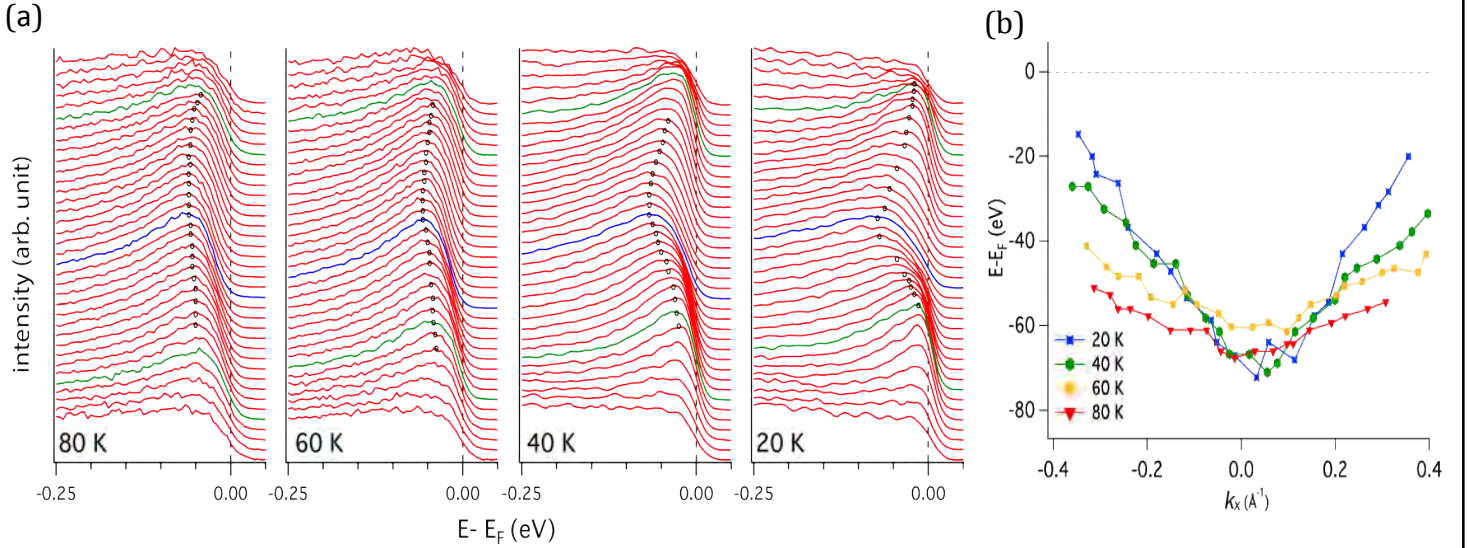

Sup. 4: (a) EDC corresponding to the cuts in Fig. (3c1-3c4). The electron-like curvature is already clear at 80 K, but the band crosses $E_F$ only at low temperatures. The blue line marks the EDC at $(\pi, \pi)$ and the green line the EDC, where the peak is closest to Γ, which we define as "$k_F$" ($k_F \cong \pm 0.38$ Å$^{-1}$ at 20 K). These spectra are reported in Fig. 4 and are used to define the pseudogap. Open circles mark the local maxima, blue lines: EDCs through $(\pi, \pi)$; green lines: EDCs through $k_F$. (b) Temperature dependent dispersions extracted by marking the local maximum of the EDCs from Fig. (3c1-3c4).

Note that this band is significantly stronger in our measurement than in ref. [22,23], where it is barely seen at the $E_F$. Remember that many different bands cross at $(\pi, \pi)$ (see Fig. 3c1), so that different experimental geometries and/or polarizations can easily pick up different contribution and give rise to quite different spectra. Our experimental conditions are optimized to select the $d_{xz}$ electron band in Fig. 3c1.

Sup. 4b shows the temperature dependent dispersion of the electron pocket at M, extracted by marking the local maximum of the EDCs from Fig. (3c1-3c4). The EDCs close to M point get very broad at low temperature, and makes it more difficult to define where is the bottom of the electron band exactly. However, it is obvious that there is no clear band shift for 80 meV as which in the hole/electron band at Γ (Fig. 3a5).

In ref. [23], only the evolution at $(\pi, \pi)$ is given, no curvature could be detected to define $k_F$. It is quite possible that, within all experimental , the evolution is similar at this point in our data. Near $E_F$, their data are dominated by a small peak that does not show any clear $k$-dependence. This peak is not as clear in our data, but it is not inconsistent, as there is a step forming near $E_F$, which could have a similar origin. Both sets of data point to a larger coherence at low temperatures.

## 5. Dispersion in $Fe_{1.09}Te_{0.78}Se_{0.22}$ :

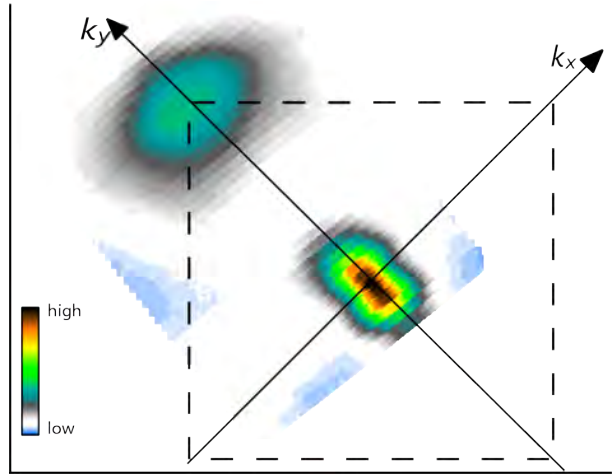

Sup. 5: FS map of $Fe_{1.09}Te_{0.78}Se_{0.22}$ at $k_z \cong 1$ integrated over $E_F \pm 5$ meV, measured at 20 K, 38 eV linear polarization (LP) along $k_x$.

For comparison, we display the FS (as shown in Sup. 5) and dispersions of on Se doped sample (as shown in Fig. 3a6, 3b6 and 3c6) at 20 K, which shows no magnetic transition.

In Fig. 3a6 and 3b6, we plot the LDA bands shifted down by 120 meV and renormalized by factor 2 for the hole pockets. For the electron pocket at M, we plot the LDA calculation that is only renormalized by factor 2 and no shift (Fig. 3c6). The different shift for fitting the electron and the hole bands can simply be understood for balance the number of hole and electron. We caution that the shift of the bands compared to the LDA calculation often changes significantly the bare Fermi velocities, as the bands are often far from linear in this energy window. This makes it more difficult to define the renormalization. Also, it would be obviously different if they were fitted independently for $d_{xz}$ (a) and $d_{yz}$ (b), questioning its general meaning. Moreover, we chose a value describing roughly the main band structure, but there could be additional renormalization near $E_F$. This may be the case for $d_{yz}$ (b) near $E_F$, although the curvature of the bare band itself could be sufficient to explain its shape.

A very curious observation is that there is no detectable $d_{xz}$ electron pocket at all for the Se doped case (Fig. 3c6). The observed band seems to fit with the saddle $d_{yz}$ band, but it normally has small intensity due to folding, as is well observed in FeTe (Fig. 3c1) and Fe- pnictides[26]. Comparison between Fig. 3c6 and Fig. 3c1 rather suggests that the observed feature is the electron band, but that it has completely opened up and even changed it curvature.